\documentclass[]{raa}            
\usepackage{graphicx,times}             
\usepackage{txfonts}	                         
\usepackage{natbib}
\begin{document}

\title{A Fast Ellipsoid Model for Asteroids Inverted From Lightcurves
$^*$
\footnotetext{$*$ Supported by Science and Technology Development Fund, MSAR.}
}
   \volnopage{Vol.0 (200x) No.0, 000--000} 
   \setcounter{page}{1}   

\author{Xiaoping Lu\inst{1} \and Haibin Zhao\inst{2} \and Zhong You\inst{1} 
   }

\institute{Faculty of Information Technology, Macau University of Science and Technology, Avenida Wai Long, Taipa, Macau, China; {\it xplu@must.edu.mo}
        \and
Purple Mountain Observatory,  Chinese Academy of
          Sciences, Nanjing, China\\
\vs \no
   {\small Received [year] [month] [day]; accepted [year] [month] [day] }
}

\abstract{
The research about asteroids attracts more and more attention recently, especially focusing on their physical structures, such as the spin axis, the rotation period and the shape. The long distance between Earth observers and asteroids makes it impossible to get the shape and other parameters of asteroids directly with the exception of the NEAs (Near Earth Asteroids) and others passed by some spacecrafts. Generally photometric measurement is still the main way to obtain the research data for asteroids now, i.e. the lightcurves recording the brightness and positions of asteroids. Supposing that the shape of the asteroid is a triaxial ellipsoid with a stable spinning status, a new method is present in this article to reconstruct the shape models of asteroids from the lightcurves, with the other physical parameters together.
By applying a special curvature function, the method calculates the brightness integration on a unit sphere and Lebedev Quadrature is employed for the discretization.    
At last the method searches the optimal solution by Levenberg-Marquardt algorithm to minimize the residual of the brightness. By adopting this method not only related physical parameters of asteroids can be obtained at a reasonable accuracy, but also a simple shape model of Ellipsoid can be generated for reconstructing more sophisticated shape model further. 
}

\authorrunning{X.-P. Lu, H.-B Zhao and Z. You} 
\titlerunning{Fast Ellipsoid Model for asteroids}  

\maketitle
\section{Introduction}
\label{sec:Intro}
With the development of the technology humans have much more ability to discover the universe now. The beginning of our solar system is still an interesting problem, which results in the research about asteroids because the asteroids reserve much information about the original formation of the solar system. Most of asteroids found now lie between Mars and Jupiter, commonly called as main belt asteroids(MBAs).
Generally MBAs are orbiting the sun with a stable status with the exception that some larger bodies pass them over and the perturbations will change their orbits. 
Due to the long distance of $2.1 \sim 3.3 $AU between the earth and most of asteroids, until now the main research data about asteroids is the photometric brightness with the position coordinates of asteroids, the earth and the sun, generally i.e. lightcurves, which are recorded by many ground-based  observatories. 

Shape models with other physical parameters such as the rotation period and spin axis, especially the albedo of asteroids are the most concerned about asteroids. The total distribution of the spin axis and rotation period of MBAs can help us research the origin and evolution of the solar system.
Commonly the rotation periods of MBAs are stable with some hours less than one day. But recently it is found that the thermal radiation from the sun can change the asteroid's rate of rotation, called as YORP effect. 
\cite{2007YORP1,2007YORP2} observed the rotation period of asteroid 54509 will double in about 600,000 years which confirms that the periods of asteroids may change in a very small amplitude within a long term.  

There are many methods to reconstruct the shape models of asteroids with the other physical parameters. \cite{1906RUSS} firstly started the research about the shape of asteroids and concluded a pessimistic result that the shape could not be obtained merely depending on the lightcurves of the opposition. With the development of both mathematical theories and technologies in telescope more and more lightcurves viewed in various solar phase angles are recorded by many observatories located everywhere on the earth and the shape inversion algorithms can figure out the shape model in a reasonable accuracy by the high-speed computers. \cite{1981LBa,1981LBb} presented an estimation about the scattering law in the surface of atmosphereless bodies and introduced a spherical harmonics method for asteroid pole determination \citep{1990LKB}. \cite{1984Hapke} moved further about the scattering law with the consideration of bidirectional reflectance and the roughness of the surface. Basing on the scattering law of Lumme and Bowell, \cite{1989Kart1,1989Kart2} presented a method to generate lightcurves of triaxial ellipsoid models and discussed the variations of convex model and nonconvex model. But they just noted that it was possible to determine the axial ratios of a triaxial ellipsoid model, while did not give a method to find the shape models from lightcurves. \cite{1989Cellino} adopted a model formed by merging together eight octants of ellipsoids having different semi-axes and generated the corresponding lightcurves without an inverse method, too. 
Adopting the sparse photometric data collected by the Hipparcos satellite,
\cite{2009Cellino} presented a genetic inversion to find the physical properties of asteroids with the assumption that the shape model of the asteroid is a triaxial ellipsoid. Further more \cite{1992MK1,1992MK2} built up a very efficient method to reconstruct the arbitrary surface of asteroids and the inverse shape models are confirmed by the flyby observation in space \citep{2002MK}. 

Nevertheless until now the ellipsoid model plays an important role. Firstly the photometric data observed by the observatories based on the earth is not enough accurate because of the atmosphere mist , the CCD heat, wrong operations and so on. Generally the error of photometric data is about 2\% and we can not expect to obtain a very accurate shape model only from the lightcurves. Secondly the ellipsoid model is simple but it can make the inversion easy and efficient with an acceptable physical parameters while the statistical research about the spin axis and period of asteroids can obtain the needed data in a fast way. Thirdly the rough model inferred from the ellipsoid model may be the initial value for reconstructing more accurate shape model by Kaasalainen's method. 

Under the similar definition employed in Kaasalainen's method and tiling the triangular facets in Lebedev way we present a fast method of the ellipsoid model in this article as the following organization. In section \ref{sec:Model} we describe the technical details of the fast ellipsoid model, including the scattering law in \ref{subsec:Scatter} and the photometric integration by applying the curvature function in \ref{subsec:Integral}. Furthermore the inverse problem with a discretized integration is presented in \ref{subsec:Inverse}. Finally the formulas of derivatives are given out for reference in optimization part \ref{subsec:Optimal}. 
At last the summing up and future plan will be discussed in section \ref{sec:Con}.

\section{Ellipsoid Model for Asteroids}
\label{sec:Model}
The disk-integrated photometric data i.e. brightness in lightcurves, contains much information about the asteroid which can be applied to obtain the related parameters. The periodic variation of brightness is mainly due to the variation of the shape as the asteroid spins around its axis. We assume that the shape of asteroid is a triaxial ellipsoid with three semi-axis $a\geq b\geq c > 0$, which spins in the period $P$ expressed in hours around its shortest axis whose spherical coordinate is denoted as ($\lambda,\beta$) in the J2000 ecliptic frame system. As the convention the brightness data recorded in lightcurves is reduced to unit distances of between the asteroid and the sun, the earth respectively and corrected according to the light-time \citep{DAMIT}. So we suppose lightcurves is processed in this way before applying our method.

\subsection{The Scattering Law}
\label{subsec:Scatter}
The scattering behavior is an inevitable problem in the asteroid model. \cite{1981LBa,1981LBb} considered several physical parameters, such as the single-scattering albedo $\Omega_0$, the asymmetry factor $g$, the volume density of the surface material $D$, the roughness of the surface $\varrho$ etc. At last a sophisticated scattering model was built to simulate the reflection behavior of the light from Sun. \cite{1984Hapke} took into account the opposition effect and the shadow in the particles of the surface. These scattering models can express the physical characteristics of the light reflection in a rational manner, but they are not efficient in reality due to the uncertain physical parameters. \cite{2005MKlab} made a photometry research about an artificial asteroid in the laboratory experiments and confirm that the shape variation is the main cause for the variation of brightness, not the scattering law. Further more, they found that it is hard to distinguish the difference between the scattering law and the random error. In order to acquire the shape model in an efficient way, a simple scattering law is needed. \cite{2001MKopti2} presented an convenient method to simulate the scattering behavior by merging both the single scattering factor $S_{LS}$(Lommel-Seeliger) and the multiple scattering factor $S_L$(Lambert) in a linear combination. The scattering law can be expressed in 
\begin{eqnarray}	
S(\mu,\mu_0,\alpha)  &=& f(\alpha)\big[ S_{LS}(\mu,\mu_0)+cS_L(\mu,\mu_0)\big]  \nonumber \\
				 &=& f(\alpha)\big(\frac{\mu\mu_0}{\mu+\mu_0}+\gamma\,\mu\mu_0\big), \label{Eqn:Sfun}
\end{eqnarray}
where the $\mu,\mu_0$ are defined as follows under the definition of $\eta(\vartheta,\varphi)$ as the outward unit normal vector of the surface and $\omega, \omega_0$ as the directions to the earth and the sun observed from the asteroid respectively,
\begin{equation}
\mu=\omega\cdot\eta,\quad \mu_0=\omega_0 \cdot \eta.
\end{equation}
The phase function $f(\alpha)$ is a fitted function in the three-parameter form 
\begin{equation}
f(\alpha)=A_0 \exp\left(-\frac{\alpha}{D}\right)+k\alpha+1,
\end{equation}
where $A_0$ and $D$ are the amplitude and scale length of the opposition effect and $k$ is the overall slope of the phase curve. The above scattering law with four parameters adopted in this article can perform efficiently in the shape inverse problem and simulate the opposition effect rationally.

\subsection{The Photometric Integration} 
\label{subsec:Integral}
\begin{figure}[ht]
\centering
\includegraphics[width=0.45\textwidth]{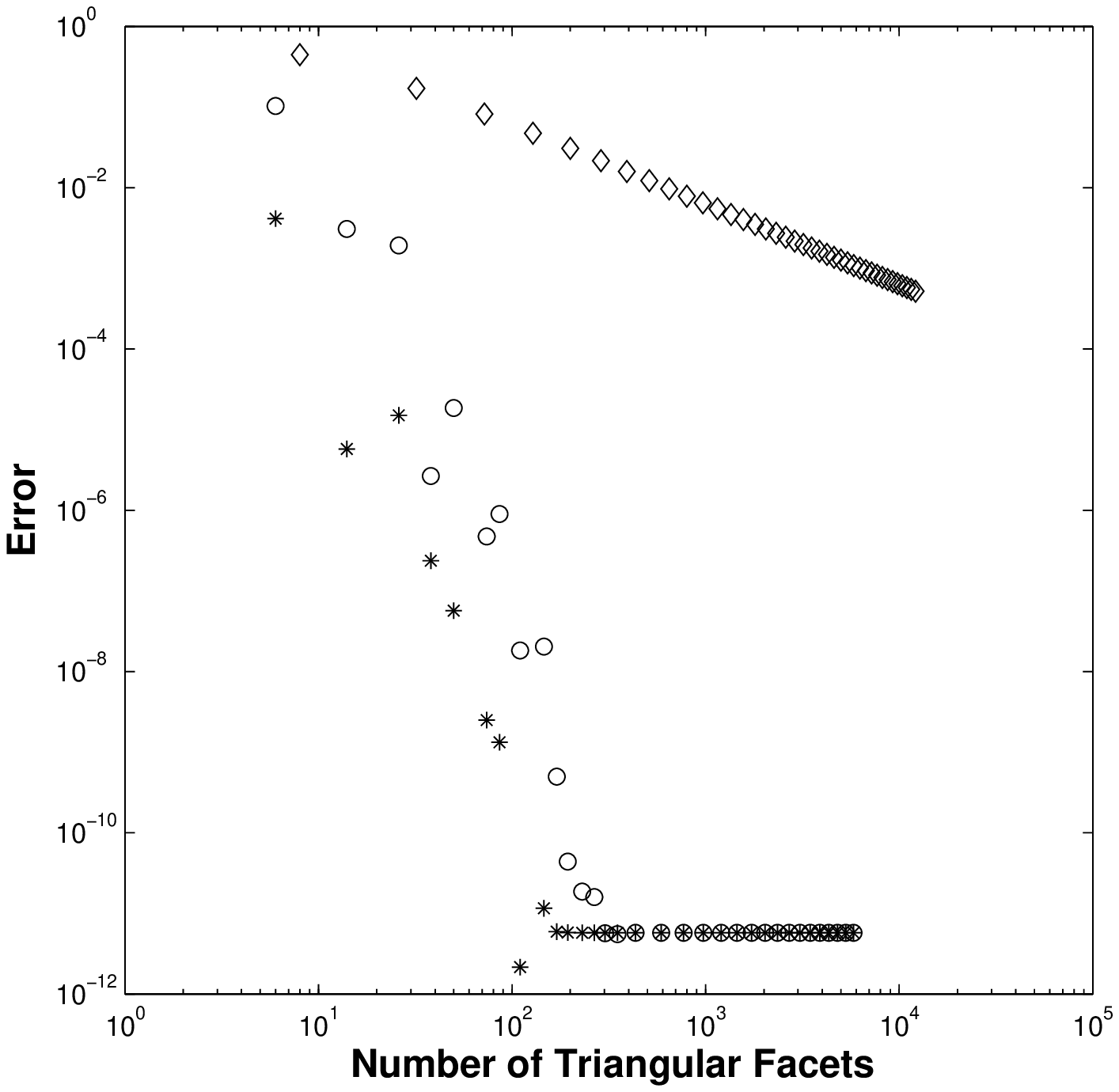}
\includegraphics[width=0.45\textwidth]{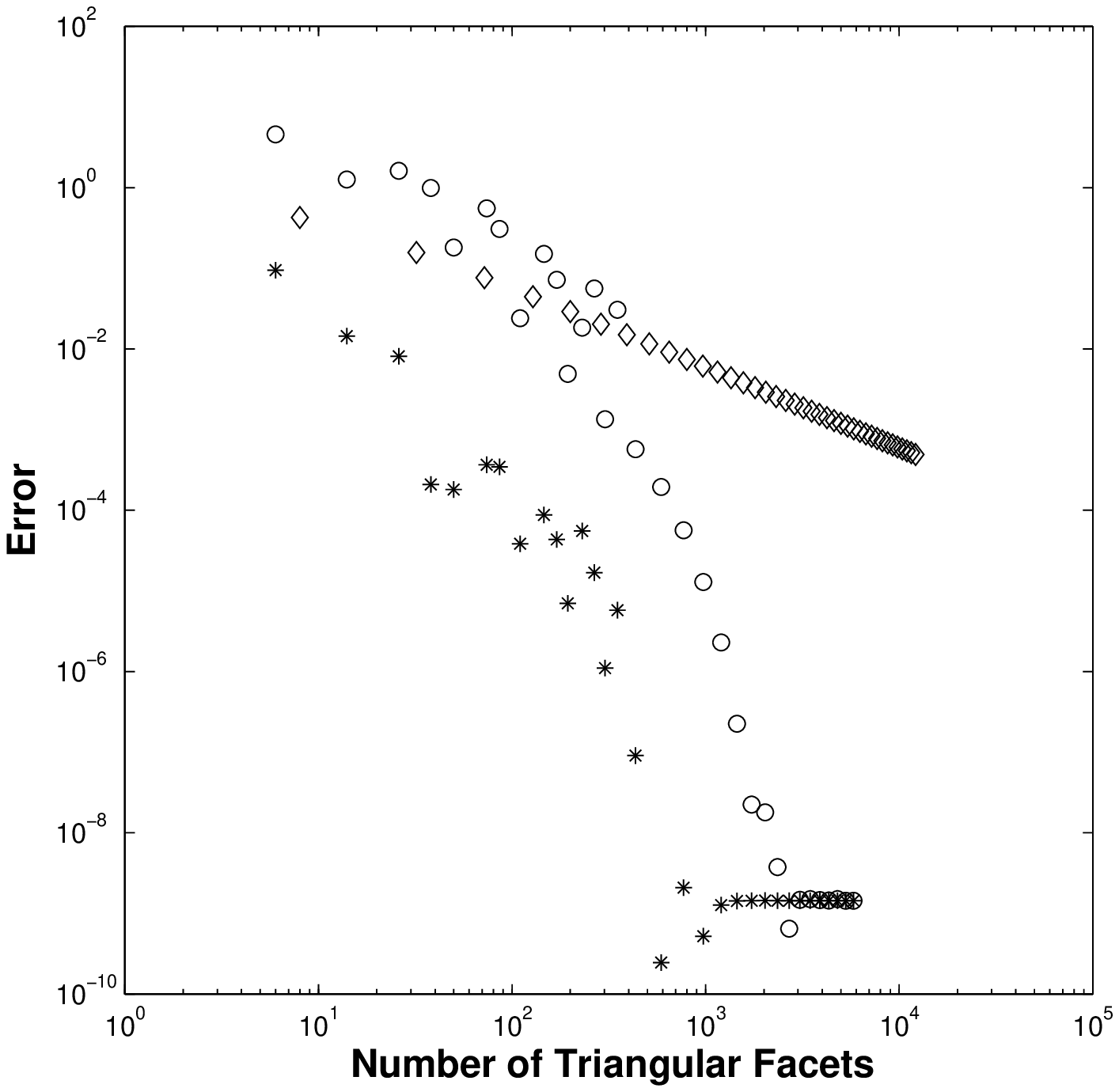}
\caption{The comparison of Triangulation(`$\diamondsuit$') and Lebedev Quadratures with our Curvature(`$*$') and Kaasalainen's Curvature(`$\circ$') respectively for computing the surface area of ellipsoids with various semi axes (Left: \textbf{a=8,b=7,c=6}  Right: \textbf{a=10,b=2,c=1.5})}
\label{Fig:2CurvVS}
\end{figure}

In order to reconstruct the shape model of asteroids, the synthetic brightness must be described in the direct problem. As mentioned above, the scattering law with four parameters can be adopted herein to generate the photometric brightness as a surface integral
\begin{equation}
\label{Eqn:Integral}
L(\omega_0,\omega)=\iint_{E_+} S(\mu,\mu_0,\alpha)\, d\sigma,
\end{equation} 
where $E_+$ is the part of the asteroid shape which is both illuminated by the sun and visible from the earth, i.e $\mu, \mu_0>0$. As assuming that the shape model of asteroids is triaxial ellipsoid, the integral (\ref{Eqn:Integral}) can be calculated numerically by the traditional triangulation, tiling the approximately equal triangular facets on the surface, which is a linear algorithm as the number of tessellated facets. In Fig \ref{Fig:2CurvVS} it is shown that an error level of $10^{-4}$ needs more than $10^4$ triangular facets tessellated on the surface of the ellipsoid.

\begin{figure}[ht]
\centering
\includegraphics[width=0.45\textwidth]{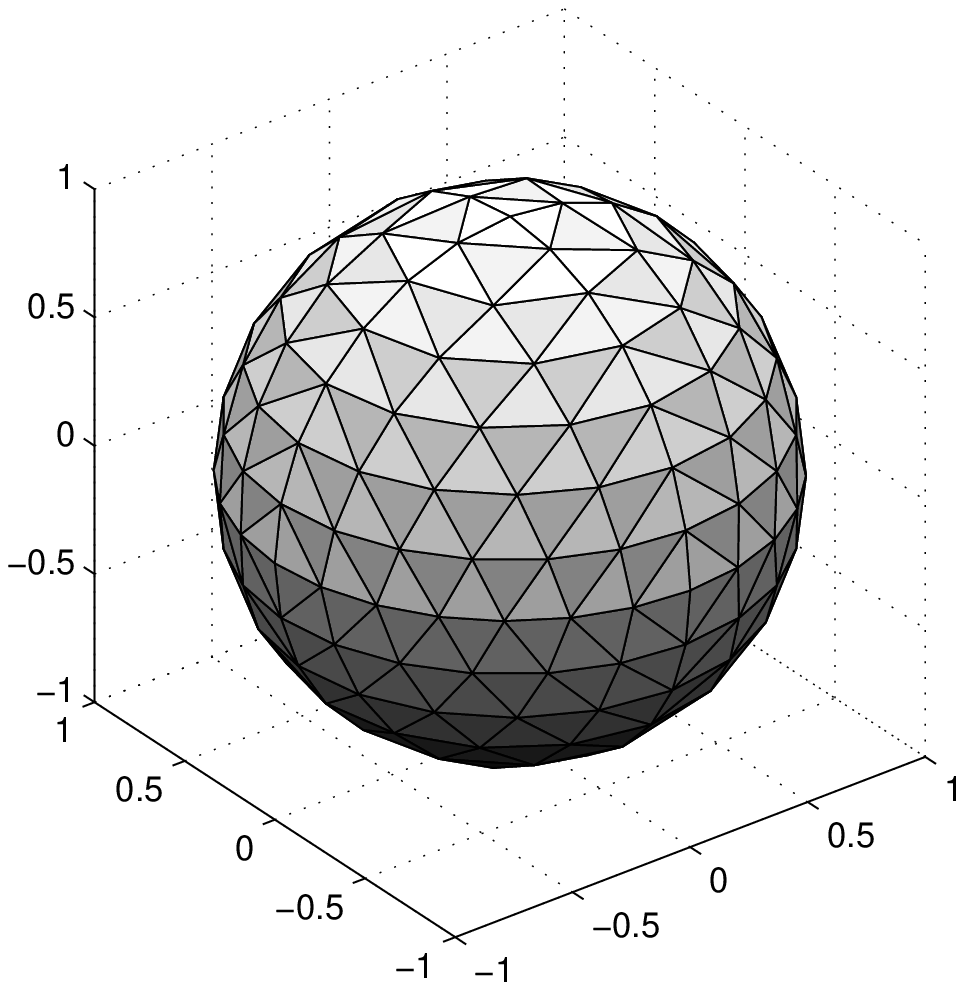}
\includegraphics[width=0.45\textwidth]{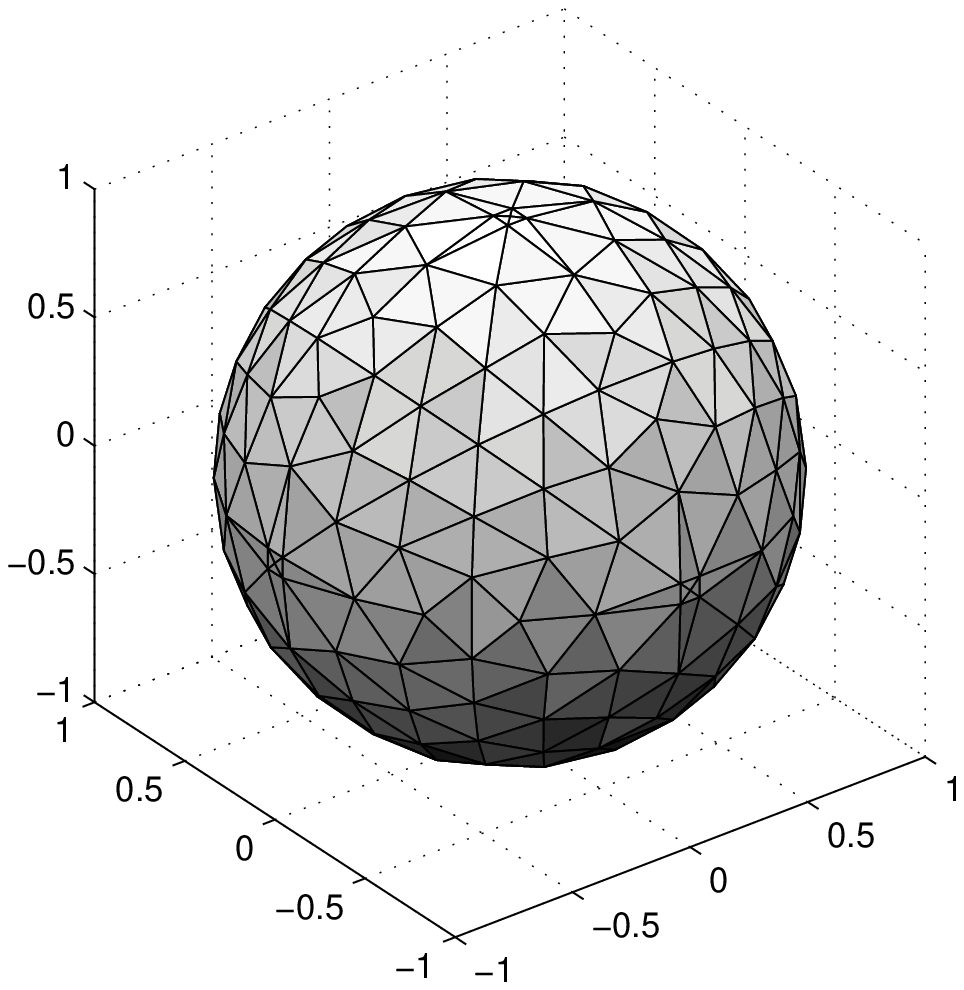}
\caption{The comparison of two different tessellation methods (Left: \textbf{Triangulation}  Right: \textbf{Lebedev})}
\label{Fig:triVS}
\end{figure}

\cite{1999Lebedev} presented a fast method to calculate the surface integral on the unit sphere $S$ by tiling the triangular facets not equally. \cite{2012luxp} applied this technique into their method and confirmed the lebedev quadrature is efficient in the surface integration. The different distributions of triangular facets of the traditional triangularization and lebedev quadrature are shown in Fig \ref{Fig:triVS}. 

Due to that the lebedev quadrature is based on the unit sphere, in our method a curvature function from the surface of ellipsoid $E$ to the unit sphere $S$ is built with the format
\begin{equation}
\label{Eqn:Curvature}
G(\theta,\phi)=\frac{\big |\frac{\partial \vec{r}}{\partial \theta} \times \frac{\partial \vec{r}}{\partial \phi}\big |}{sin\theta}, \vec{r}(\theta,\phi) \in E.
\end{equation}
With the curvature function $G(\theta,\phi)$ the brightness integral (\ref{Eqn:Integral}) can be transformed to the surface integral on the unit sphere $S$
\begin{equation}
\label{Eqn:IntegralS}
L(\omega_0,\omega)=\iint_{S_+} S(\mu,\mu_0,\alpha)\, G(\theta, \phi) \, d\sigma,
\end{equation}
where $S_+$ is the similar part of unit sphere with $\mu,\mu_0>0$.

The curvature function under the standard parametrization of the ellipsoidal surface $E$ 
\begin{equation}
\label{Eqn:Para}
x_1=a\sin\theta\cos\phi, x_2=b\sin\theta\sin\phi, x_3=c\cos\theta,  \theta \in [0,2\pi], \phi \in [0,\pi],
\end{equation}	
has the form
\begin{equation}
\label{Eqn:CurvLU}
G(\theta,\phi)=abc\sqrt{\left(\frac{\sin\theta\cos\phi}{a}\right )^2+\left (\frac{\sin\theta\sin\phi}{b}\right )^2+\left (\frac{\cos\theta}{c}\right)^2},
\end{equation}
while \cite{1992MK2} also presented the other curvature function with the form
\begin{equation}
\label{Eqn:CurvMK}
G(\vartheta,\varphi)=\left ( \frac{abc}{(a\sin\vartheta\cos\varphi)^2+(b\sin\vartheta\sin\varphi)^2+(c\cos\vartheta)^2} \right )^2
\end{equation}
under the definition of $(\vartheta,\varphi)$ denoting the spherical coordinates of the normal vectors of the ellipsoidal surface. 

Both of two curvature functions can perform well in computing the brightness integral (\ref{Eqn:IntegralS}). But the first one mentioned and adopted in our method is more efficient for the extreme situations, such as the ellipsoid model with a large difference between its three semi axes. The `Dog-Bone' shaped asteroid (216) Kleopatra has been found that its shape is close to an elongated ellipsoid \citep{2011Descamps}. Fig \ref{Fig:2CurvVS} shows the comparison of our curvature function(`$*$') and Kaasalainen's curvature function(`$\circ$') to compute the surface area of ellipsoids with different semi axes with the following formula easily letting the scattering function $S(\mu,\mu_0,\alpha)=1$ in the integral \ref{Eqn:IntegralS},
\begin{equation}
	Surface\hspace{0.5em} Area \hspace{0.5em} of \hspace{0.5em} Ellipsoids = \iint_{S_+} G(\theta, \phi) \, d\sigma.
\end{equation}	
Apparently our curvature can obtain lower error lever than Kaasalainen's curvature, especially in the case of elongated ellipsoids. Besides, the difference between the triangulation(`$\diamondsuit$') and lebedev quadrature(`$*$' or `$\circ$') is also compared in Fig \ref{Fig:2CurvVS}. The dominant lebedev quadrature can guarantee the efficiency of our fast method. 

\subsection{The Inverse Problem}
\label{subsec:Inverse}
With the definitions of both the scattering function and the brightness integration the inverse problem of the shape model of asteroids can be illustrated totally as follows.

Supposing that all lightcurves in this article are formatted in the same way of DAMIT \citep{DAMIT}, the relative brightness ($L$) with respect to a Julian Day ($t$) and the positions of the earth and sun ($\omega,\omega_0$) in ecliptic asteroid-centric coordinate system can be obtained directly from lightcurves. According to the aforementioned brightness integration, all required parameters are three semi axes ($a,b,c$), the spherical coordinate ($\lambda,\beta$) of the spin axis in ecliptic system, the period ($P$), the four parameters ($A_0,D,k,\gamma$) of the scattering function and the initial phase angle ($\Phi_0$) at the beginning epoch ($t_0$). So there are 11 parameters in all as $t_0$ is generally set to be the first Julian date of all lightcurves. 

The observed brightness will vary periodically as the rotation of the asteroid around its spin axis, which seems that the sun and earth spin around the stationary asteroid. Assuming the asteroid coordinate system is the Cartesian frame with the spin axis as z-axis and the long semi-axis as x-axis, the origins of ecliptic and asteroid coordinate system are located at the same point of the center of the asteroid. Denoting $\tilde{\omega},\tilde{\omega_0}$ as the coordinate of the earth and sun in the asteroid coordinate system, the transformation between ecliptic system and asteroid system can be given as
\begin{equation}
	\tilde{\omega}=R_3(\Phi)R_2(\beta)R_3(\lambda)\omega, \hspace{2em}				   \tilde{\omega_0}=R_3(\Phi)R_2(\beta)R_3(\lambda)\omega_0,
\end{equation}
where $R_2,R_3$ are the rotation matrices rotated as the y and z axis respectively with the forms
\begin{equation}
	R_2(\theta)=\left( \begin{array}{ccc}
	\cos\theta & 0 & -\sin \theta \\
	0 & 1 & 0 \\
	\sin \theta & 0 & \cos \theta \end{array} \right),  
	R_3(\theta)=\left( \begin{array}{ccc}
	\cos\theta & \sin\theta & 0 \\
	-\sin\theta & \cos\theta & 0 \\
	0 & 0 & 1 \end{array} \right).	
\end{equation}
As the asteroid rotates in an angular speed $2\pi/P$, phase angle $\Phi$ varies with respect to the epoch $t$ (Julian Day) in the form
\begin{equation}
	\label{Eqn:PhaseFun}
	\Phi(t)=\Phi_0+\frac{2\pi}{P}(t-t_0),
\end{equation}
while if the YORP effect is taken into account, the corresponding form will be
\begin{equation}
	\Phi(t)=\Phi_0+\frac{2\pi}{P}(t-t_0)+\frac{1}{2}v(t-t_0)^2,
\end{equation}
where $v=d\Omega/d t ~(\Omega=2 \pi/P)$ is the rotation rate of change of the period $P$. 

Let ($\theta_i,\phi_i$) be the discretized points of Lebedev quadrature with the total number of facets $N$, tessellating on the surface of unit sphere with a corresponding weight $w_i$, which can be also treated as the area of small facets. With the normal vector of any small facet of the ellipsoid shape 
\begin{equation}
	\label{Eqn:NormalVec}
	\vec{\eta}_i=\left ( \frac{\sin\theta_i\cos\phi_i}{a}, \frac{\sin\theta_i\sin\phi_i}{b}, \frac{\cos\theta_i}{c}\right ),
\end{equation} 
the brightness integral (\ref{Eqn:IntegralS}) can be discretized as 
\begin{equation}
\label{Eqn:IntegralD}
L(\omega,\omega_0) \approx \sum^N_{i=1}\left ( S(\mu^{(i)},\mu^{(i)}_0,\alpha)G(\theta_i,\phi_i)w_i\right),	
\end{equation}
where $\mu^{(i)},\mu^{(i)}_0$ are the inner products of $\tilde{\omega},\tilde{\omega_0}$ and the unit normal vector $\vec{\eta_i}/|\vec{\eta_i}|$ respectively. Further more under the definition of normal vector (\ref{Eqn:NormalVec}), the curvature function (\ref{Eqn:CurvLU}) can be simplified in the form
\begin{equation}
G(\theta_i,\phi_i)=abc|\vec{\eta_i}|.
\end{equation}
Finally merging the scattering function $S(\mu,\mu_0,\alpha)$ in the formula (\ref{Eqn:Sfun}) the discretized brightness integral (\ref{Eqn:IntegralD}) will be
\begin{equation}
	\label{Eqn:IntegralSimple}
L(\omega,\omega_0) \approx \sum^N_{i=1}\left (f(\alpha)(\vec{\tilde{\omega}} \cdot \vec{\eta_i})(\vec{\tilde{\omega_0}} \cdot \vec{\eta_i})\left [ \frac{1}{(\vec{\tilde{\omega}} \cdot \vec{\eta_i})+(\vec{\tilde{\omega_0}} \cdot \vec{\eta_i})}+\frac{\gamma}{|\vec{\eta_i}|}\right]a b c w_i \right),	
\end{equation}
where $(\vec{x}\cdot \vec{y})$ is the inner product of two vectors $\vec{x}, \vec{y}$.

The inverse problem of the ellipsoid model can be described to find the 11 parameters mentioned above to minimize the 
\begin{equation}
	\label{Eqn:IP}
\chi^2=\sum_i\left \| \frac{\vec{L}^{(i)}}{\langle \vec{L}^{(i)}\rangle}-\frac{\vec{\tilde{L}}^{(i)}}{\langle \vec{\tilde{L}}^{(i)}\rangle} \right \|^2,
\end{equation}
where $\vec{L}^{(i)},\vec{\tilde{L}}^{(i)}$ denote the brightness vectors of observed data and the synthetic data containing all points in the $i$th lightcurve, while $\langle \vec{L}^{(i)}\rangle,\langle \vec{\tilde{L}}^{(i)}\rangle$ denote the mean brightness of the two brightness vectors.

\subsection{The Optimization Algorithm}
\label{subsec:Optimal}
There are many methods to find the best fit solution of the inverse problem (\ref{Eqn:IP}), such as the genetic algorithm, methods of conjugate gradients and so on. Herein we employ the Levenberg-Marquardt method to search the optimal solution, which works very well in practice and has almost become the standard of nonlinear least-squares routines \citep{1963LM}. The LM(Levenberg-Marquardt) method can converge in a very fast speed, while the obtained result is often the local optimal solution. There is a general method to complement its deficiency, i.e. searching the best fit result with various initial values. The details of LM method can be found in \cite{1992NR}. 

But it needs to compute the derivatives of $\chi^2$ in LM method, which means the derivatives of the discretized brightness integral (\ref{Eqn:IntegralSimple}) with respect to the 11 parameters have to be computed in formula! In order to make it program easily we give out the formulas for reference.  

Letting 
\begin{equation}
	\tilde{\mu_i}=\tilde{\vec{\omega}}\cdot\vec{\eta_i},\tilde{\mu_i}=\tilde{\vec{\omega}}\cdot\vec{\eta_i},
	S=\frac{\tilde{\mu}\tilde{\mu_0}}{\tilde{\mu}+\tilde{\mu_0}}+\frac{\gamma\tilde{\mu}\tilde{\mu_0}}{|\vec{\eta_i}|},
\end{equation}	
the derivative of $L(\omega,\omega_0)$ with respect to long semi axis($a$) has the form 
\begin{equation}
	\frac{\partial L}{\partial a}=\sum_{i=1}^N \left [ f(\alpha)b c w_i S+abc w_i \left (\frac{\partial S}{\partial \mu}\frac{\partial \mu}{\partial a}+\frac{\partial S}{\partial \mu_0}\frac{\partial \mu_0}{\partial a}+\frac{\gamma \mu \mu_0 \sin^2\theta_i \cos^2\phi_i}{a^3|\vec{\eta_i}|^3}\right ) \right ].
\end{equation}
The other semi axes (such as $b,c$) have the similar formulas. The derivatives of $L(\omega,\omega_0)$ with respect to $(\lambda,\beta,P,\Phi_0)$ are
\begin{equation}
	\frac{\partial L}{\partial \lambda}=\sum_{i=1}^N \left [ f(\alpha)a b c w_i \left (\frac{\partial S}{\partial \mu} R_3(\Phi)R_2(\beta)\frac{\partial R_3(\lambda)}{\partial \lambda}\vec{\omega}\cdot \vec{\eta_i}+\frac{\partial S}{\partial \mu_0}R_3(\Phi)R_2(\beta)\frac{\partial R_3(\lambda)}{\partial \lambda}\vec{\omega_0}\cdot \vec{\eta_i} \right ) \right ],
\end{equation}	

\begin{equation}
	\frac{\partial L}{\partial \beta}=\sum_{i=1}^N \left [ f(\alpha)a b c w_i \left (\frac{\partial S}{\partial \mu} R_3(\Phi)\frac{\partial R_2(\beta)}{\partial \beta} R_3(\lambda)\vec{\omega}\cdot \vec{\eta_i}+\frac{\partial S}{\partial \mu_0}R_3(\Phi)\frac{\partial R_2(\beta)}{\partial \beta} R_3(\lambda)\vec{\omega_0}\cdot \vec{\eta_i} \right ) \right ],
\end{equation}

\begin{equation}
	\frac{\partial L}{\partial \Phi_0}=\sum_{i=1}^N \left [ f(\alpha)a b c w_i \left (\frac{\partial S}{\partial \mu} \frac{\partial R_3(\Phi)}{\partial \Phi}R_2(\beta)R_3(\lambda)\vec{\omega}\cdot \vec{\eta_i}+\frac{\partial S}{\partial \mu_0}\frac{\partial R_3(\Phi)}{\partial \Phi}R_2(\beta)R_3(\lambda)\vec{\omega_0}\cdot \vec{\eta_i} \right ) \right ].
\end{equation}
Due to the relation of $P$ and $\Phi_0$ in one function (\ref{Eqn:PhaseFun}), 
\begin{equation}
	\frac{\partial L}{\partial P}=	\frac{\partial L}{\partial \Phi_0}\frac{\partial \Phi}{\partial P}.	
\end{equation}	
The derivatives of $L(\omega,\omega_0)$ with respect to the three parameters $A_0,D,k$ in scattering function are easy to obtain by multiplying the corresponding derivatives of $f(\alpha)$ respectively,
\begin{equation}
	\frac{\partial f}{\partial A_0}=\exp(-\frac{\alpha}{D}),\hspace{1em} \frac{\partial f}{\partial D}=\frac{A_0\alpha}{D^3}\exp(-\frac{\alpha}{D}),\hspace{1em} \frac{\partial f}{\partial k}=\alpha, 
\end{equation}	
and the derivative to $\gamma$ is
\begin{equation}
	 \frac{\partial L}{\partial \gamma}=\sum_{i=1}^N \left [ f(\alpha)a b c w_i \frac{\mu\mu_0}{|\vec{\eta_i}|} \right ].
\end{equation}

\section{Conclusions}
\label{sec:Con}
We have described a fast method to obtain the physical parameters and the shape models of asteroids basing on the ellipsoidal shape. This method adopted the Lebedev quadrature to discretize the surface integral on the unit sphere, which can decrease the computational cost largely with a high accuracy. In addition now that this method can compute the period and the orientation of the spin axis in an efficient way, we can rely on its result to refine the shape of asteroids, such as adopting Kaasalainen's method. 

The related formulas are presented in this article totally, such as curvature function (\ref{Eqn:CurvLU}) and  discretized brightness integral (\ref{Eqn:IntegralSimple}), especially the complicated derivatives of $\chi^2$. Here it should be noted that all the formulas are given out in a general form with three semi-axis parameters $a, b, c$ because this format is comprehensible and easy to compare with other formula like Kaasalainen's Curvature function. In computation due to the relative brightness the semi-axis parameters $a, b, c$ are not exactly same as the real asteroids in size. So the ratios of axis of ellipsoid model $a/b, a/c$ are more practical. By applying the fast method, it is easy to compute the ratios from the derived parameters at last or we can set $c=1$ in Levenberg-Marquardt method to obtain the ratios directly. Besides, we can also replace the $\frac{2\pi}{P}$ in $\Phi(t)$ by its angular speed $\Omega$ to simplify the computation. The detailed numerical test and application to real asteroids will be presented in the future article.
  
\begin{acknowledgements}
The authors thank the anonymous referee for comments. This work is funded by the grant No. 019/2010/A2 from Science and Technology Development Fund, MSAR.  Haibin Zhao thanks	the support of the National Natural Science Foundation of China (Grant Nos. 10503013, 11078006 and 10933004) and the Minor Planet Foundation of Purple Mountain Observatory. Besides, We sincerely appreciate Prof. Kaasalainen's constructive advices about this fast method.
\end{acknowledgements}

\end{document}